# The Normalization of Occurrence and Co-occurrence Matrices in Bibliometrics using *Cosine* Similarities and *Ochiai* Coefficients

Qiuju Zhou [a] & Loet Leydesdorff *[b]

[a] National Science Library, Chinese Academy of Sciences, 100190, Beijing, People's Republic of China; email: zhouqj@mail.las.ac.cn

[b] ***corresponding author; University of Amsterdam, Amsterdam School of Communication Research (ASCoR), PO Box 15793, 1001 NG Amsterdam, The Netherlands; email: loet@leydesdorff.net

## Abstract

We prove that Ochiai similarity of the co-occurrence matrix is equal to cosine similarity in the underlying occurrence matrix. Neither the cosine nor the Pearson correlation should be used for the normalization of co-occurrence matrices because the similarity is then normalized twice, and therefore over-estimated; the Ochiai coefficient can be used instead. Results are shown using a small matrix (5 cases, 4 variables) for didactic reasons, and also Ahlgren *et al.*'s (2003) co-occurrence matrix of 24 authors in library and information sciences. The over-estimation is shown numerically and will be illustrated using multidimensional scaling and cluster dendograms. If the occurrence matrix is not available (such as in internet research or author co-citation analysis) using Ochiai for the normalization is preferable to using the cosine.

**Keywords**: normalization, occurrence, co-occurrence, affiliation, Ochiai, cosine, overlap

## Introduction

Ahlgren *et al.* (2003) argued that in the case of bibliometric co-occurrence data, the use of the Pearson correlation coefficient *r* is problematic: two natural requirements of a similarity measure applied, for example, in author cocitation analysis are not satisfied by *r*. However, an alternative is provided by using the cosine. Using Salton's cosine similarity instead of the Pearson correlation coefficient for the normalization addresses two problems (*i*) the skewness of the distribution in bibliometric data (Seglen, 1992) and (*ii*) the expected prevalence of zeros in most of the vectors of the citation matrix.

The cosine similarity is equal to the Pearson correlation coefficient except that the cosine is not normalized with reference to the mean of the distribution, while the Pearson correlation is. The cosine similarity can therefore be considered a non-parametric measure. Egghe & Leydesdorff (2009) showed that the correspondence between these two measures (cosine and Pearson) is not linear, but can be represented as a sheaf of straight lines. Note that the Pearson correlation also implies *z*-normalization of the variation, whereas the cosine does not.

The argument of Ahlgren *et al.* (2003) led to an intensive debate in this journal (Ahlgren *et al*., 2004; Bensman, 2004; Leydesdorff, 2005; White, 2003 and 2004) because in bibliometrics, author cocitation analysis (ACA) had previously been based on using Pearson correlations and factor analysis (McCain, 1990; White & Griffith, 1981; White & McCain, 1998). Multi-dimensional scaling (MDS), however, is also non-parametric and can therefore be based on cosine-normalized matrices. Leydesdorff & Vaughan (2006) argued that one should not normalize the co-occurrence matrix using the Pearson correlation or cosine, but use the underlying occurrence (e.g., word-document) matrix for the normalization instead of the co-occurrence matrix. The co-occurrence matrix—co-citation, co-word, co-authorship, etc., matrix—can be derived from the occurrence matrix through multiplication by its transpose. But one cannot derive the occurrence matrix from the co-occurrence matrix because information is lost in the transformation (Leydesdorff, 1989). The co-occurrence matrix contains the inner products of the vectors that are also the numerators of the respective cosines, and thus provide a first step in the normalization.

In social network analysis, the use of the co-occurrence or affiliations matrix is common and implemented in the software (such as in Pajek and UCInet) since one is more interested in the relations between variables (e.g., co-words) and their network properties than in the attribution of variables to cases (e.g., documents). The affiliations matrix of co-occurrences provides direct access to the network.

Ahlgren *et al.* (2003) provided as an empirical example, the author co-citation matrix among 12 bibliometricians and 12 authors from the information retrieval field, and normalized this matrix using both the Pearson correlation and the cosine similarity. Leydesdorff & Vaughan (2006) reproduced this matrix and its underlying asymmetrical matrix of occurrences in order to show the differences in distinguishing between the two groups in these matrices using MDS and a spring-embedded algorithm (Kamada & Kawai, 1989). These authors suggested that whenever the asymmetrical occurrence matrix is unavailable, as in most Internet research, one should perhaps better use the Jaccard index; but the issue remained analytically unresolved. Leydesdorff (2008) compared a large number of possible indices using these same occurrence and co-occurrence matrices (cf. Jones & Furnas, 1987; Schneider & Borlund, 2007a; Van Eck & Waltman, 2009).

In summary, two problems can be distinguished: (*i*) the use of the cosine similarity versus the Pearson correlation in the case of skewed bibliometric distributions, and (*ii*) using the occurrence or co-occurrence matrix as input to the normalization. Ahlgren *et al.* (2003) provide convincing arguments for using the cosine instead of the Pearson correlation, but used the co-occurrence matrix for making their empirical argument. Leydesdorff & Vaughan (2006) argued in favour of using the asymmetrical occurrence matrix for the normalization, since the co-occurrence matrix is already normalized—providing the numerators of the cosine or, in other words, the inner products between the vectors.

In the following section we address a third source of possible confusion: the difference between cosine similarity and the Ochiai coefficient in the case of a non-binary matrix. The Ochiai coefficient can be considered as the binary variant of the cosine (Schneider & Borlund, 2007b, at p. 1599). Thereafter, we turn first to a small matrix for didactic purposes and then apply the resulting insights to the matrix that was introduced by Ahlgren *et al*. (2003) and replicated by Leydesdorff & Vaughan (2008) in making their respective arguments.

## Cosine similarity *versus* the Ochiai coefficient

Salton & McGill (1983, at p. 121; Sen & Gan, 1983, at p. 80) introduced the *cosine* between two vectors *x* and *y* into the information sciences. The cosine can be formulated as follows:

$$\text{Cosine}(x,y) = \frac{\sum_{i=1}^{n} x_i y_i}{\sqrt{\sum_{i=1}^{n} x_i^2 * \sum_{i=1}^{n} y_i^2}} \qquad (1)$$

Note that the formula of the cosine is identical to the one of the Pearson correlation, but without the centering of the vectors to the mean (Egghe & Leydesdorff, 2009).

For a *binary* matrix, Eq. 1 can be simplified as follows:

$$\text{Cosine}(x,y)^{\text{binary}} = \frac{\sum_{i=1}^{n} x_i y_i}{\sqrt{\sum_{i=1}^{n} x_i \ * \sum_{i=1}^{n} y_i}} \qquad (2)$$

since the squared norm of the vector ($L_2 = \sum_i {x_i}^2$) is equal to the sum ($L_1 = \sum_i x_i$) in the binary case.

The similarity measure in Eq. 2 is a variant of the so-called Ochiai coefficient (Driver & Kroeber, 1932, at pp. 217-219; Ochiai, 1957; cf. Bolton, 1991, at pp. 143-145; Cui.1995; Yang, 2007, at p.47 ):):

$$Ochiai(\mathrm{x},\mathrm{y}) = \frac{c_{\mathrm{xy}}}{\sqrt{c_{\mathrm{x}}\ c_y}} \qquad (3)$$

In Eq. 3, $c_x$ denotes the sum of the number of occurrences (count) of *x* and $c_{xy}$ the sum of the co-occurrences of *x* and *y*. The Ochiai coefficient is defined at the nominal scale and does not take the ordinal nature of bibliometric data into account. In the subroutine Proximities of SPSS, for example, Ochiai can be used only for binary matrices, whereas SPSS suggests using the cosine or the Pearson correlation for the non-binary case. However, SPSS rejects non-binary values when one asks for the Ochiai coefficient.[1]

---

[1] SPSS provides the formula for the Ochiai coefficient between two variables *x* and *y* as follows:

$$Ochiai(x,y) = \frac{\mathrm{a}}{\sqrt{\mathrm{a+b}}\sqrt{\mathrm{a+c}}} \qquad (4)$$

using the following 2×2 contingency table:

| | | variable *x* |
|---|---|---|

One can use Eq. 3 also as a formula for non-binary matrices.[2] Glänzel & Czerwon (1995; 1996, at p. 199) suggested using the Ochiai for a numerical co-occurrence matrix as "a simplified cosine" (Zhou *et al*., 2009, at p. 602). The use of this alternative for the cosine has led to possible confusion in the literature, as if two different definitions of the cosine were available (Van Eck & Waltman, 2009, at p. 1637 and 1645, note 9). Small & Sweeney (1985, at p. 397) used Eq. 3 for normalizing a non-binary co-citation matrix, but called it Salton's cosine similarity.

We shall show the differences between the cosine and the Ochiai coefficient using an example. But we argue that the various measures can meaningfully be used for different purposes: the Ochiai coefficient of the co-occurrence matrix is equal to the cosine of the occurrence matrix, and thus enables us to normalize the co-occurrence matrix as precisely as the (potentially absent) occurrence matrix. The Ochiai coefficient is also the best approximation of the cosine similarity in the occurrence

---

| | | | |
|---|---|---|---|
| variable *y* | Presence | a | b |
| | Absence | c | d |

[2] Jones & Furnas (1987, at pp. 429f.) propose the "pseudo-cosine" that is formalized as follows:

$$\text{Pseudo Cosine}(x, y) = \frac{\sum_{i=1}^{n} x_i y_i}{\sum_{i=1}^{n} x_i * \sum_{i=1}^{n} y_i} \qquad (5)$$

Unlike the Ochiai, the denominator is not square-rooted and therefore much larger. Consequently, the values of the pseudo-cosine are much smaller than those of the cosine.

matrix if the latter is not available; for example, when the co-occurrence matrix can be measured empirically.

### The derivation of the co-occurrence matrix from the occurrence matrix

As noted, one can derive a co-occurrence matrix from the occurrence matrix by multiplying the latter by its transposed: $\mathbf{A}^\mathbf{T} * \mathbf{A}$. Note that $\mathbf{A} * \mathbf{A}^\mathbf{T}$ provides a second co-occurrence matrix along the other dimension of the cases of the matrix. The off-diagonal values in the symmetrical co-occurrence matrix are the sums of the inner products between the vectors $(\vec{x}_i * \vec{y}_i)$, and the diagonal value is equal to the squared norm of each vector in the occurrence matrix: $|\vec{X}| * |\vec{X}|$.

Let us demonstrate this using the small (numerical) matrix of five documents and three variables (e.g., words) in Table 1:

**Table 1**: asymmetrical occurrence matrix

| | V1 | V2 | V3 |
|---|---|---|---|
| D1 | 2 | 0 | 2 |
| D2 | 1 | 1 | 0 |
| D3 | 0 | 3 | 3 |
| D4 | 0 | 2 | 2 |
| D5 | 0 | 0 | 1 |

When multiplied by its transposed (that is, after swapping rows and columns), the resulting co-occurrence matrix is provided in Table 2:

**Table 2**: symmetrical co-occurrence matrix (over the columns)

| | V1 | V2 | V3 |
|---|---|---|---|
| V1 | 5 | 1 | 4 |
| V2 | 1 | 14 | 13 |
| V3 | 4 | 13 | 18 |

V2 and V3, for example, occur both three times in document D3 and twice in D4. The cell (V2, V3) thus has a value of 3*3 + 2*2 = 13. The diagonal value, however, is based on the matrix multiplication and therefore the square of the vector. In the case of V3, for example, this value is along the column of V3 (in Table 1): 2*2 + 0*0 + 3*3 + 2*2 + 1*1 = 18.

UCINet, for example, does this matrix multiplication correctly when one asks for Affiliations in the Data menu; Pajek, however, omits the diagonal values when the 2-mode matrix of Table 1 is transformed into a 1-mode matrix; one first has to turn on the option "include loops." Alternatively, one can transpose the 2-mode matrix and then use the subroutine Networks for the multiplication of the matrices (de Nooy *et al.*, 2011). In Excel, one can use the functions TRANSPOSE() and MMULT() consecutively to generate Table 2 from Table 1.

Morris (2005, at p. 22) notes that in empirical research the co-occurrence matrix is often based on the minimal overlap between the vectors for each case, and not on matrix multiplication. While one can assume that the underlying occurrence matrix is binary in the case of co-citation or co-author matrices, linguistic term occurrence matrices are not binary since each term may occur multiple times in a paper (Morris, 2005, p. 36). The results of matrix multiplication with the transposed sometimes provide less meaningful representations in this case.

If one searches—for example, at the internet—for "a AND b", one retrieves the minimum overlap and not the multiple. The minimum overlap is in this case binary: the retrieved sets overlap or not. Using Morris (2005) non-binary overlap function between the vectors, the minimum overlap between V1 and V3 in Table 1 is 2. Table 3 provides the co-occurrence matrix based on this overlap applied to Table 1. Note that the diagonal values are now equal to the $L_1$ (= $\sum_i x_i$) norms of the respective vectors in Table 1.

**Table 3**: Symmetrical co-occurrence matrix based on Table 1, but using the minimal overlap

| | V1 | V2 | V3 |
|---|---|---|---|
| V1 | 3 | 1 | 2 |
| V2 | 1 | 6 | 5 |
| V3 | 2 | 5 | 8 |

The Ochiai coefficients based on the minimum overlap function can be formalized as follows:

$$Ochiai(\mathrm{x}, \mathrm{y}) = \frac{\sum_{i=1}^{n} \min(x_i, y_i)}{\sqrt{\sum_{i=1}^{n} x_i \quad \sum_{i=1}^{n} y_i}} \qquad (6)$$

The co-occurrence (that is the inner product) in the numerator is replaced with the minimum value for *x* AND *y*.

Let us cross-table the options of using the cosine similarity (Eq. 1) and the Ochiai coefficient (Eq. 3) for both the asymmetrical and symmetrical matrices. The result is shown Table 4, as follows:

**Table 4**: Cosine and Ochiai values for occurrence and co-occurrence matrices

| | Cosine (Eq. 1) | | | | Ochiai (Eq. 3) | | | |
|---|---|---|---|---|---|---|---|---|
| Occurrence matrix (Table 1) | | V1 | V2 | V3 | | V1 | V2 | V3 |
| | V1 | 1.00 | 0.12 | 0.42 | V1 | 1.00 | 0.24 | 0.82 |
| | V2 | 0.12 | 1.00 | 0.82 | V2 | 0.24 | 1.00 | 1.88 |
| | V3 | 0.42 | 0.82 | 1.00 | V3 | 0.82 | 1.88 | 1.00 |
| Co-occurrence matrix based on inner products (Table 2) | | V1 | V2 | V3 | | V1 | V2 | V3 |
| | V1 | 1.00 | 0.57 | 0.72 | V1 | 1.00 | 0.12 | 0.42 |
| | V2 | 0.57 | 1.00 | 0.97 | V2 | 0.12 | 1.00 | 0.82 |
| | V3 | 0.72 | 0.97 | 1.00 | V3 | 0.42 | 0.82 | 1.00 |
| Co-occurrence matrix based on overlap function (Table 3) | | V1 | V2 | V3 | | V1 | V2 | V3 |
| | V1 | 1.00 | 0.65 | 0.75 | V1 | 1.00 | 0.24 | 0.41 |
| | V2 | 0.65 | 1.00 | 0.95 | V2 | 0.24 | 1.00 | 0.72 |
| | V3 | 0.75 | 0.95 | 1.00 | V3 | 0.41 | 0.72 | 1.00 |

Table 4 shows that the cosine values of the occurrence matrix (Table 1) are precisely equal to the Ochiai values of the co-occurrence matrix (Table 2). The Ochiai coefficient of the co-occurrence matrix uses the inner products in the numerator, and the diagonal values in Table 2 (that are equal to the squared norm of the original vectors) in the denominator. Cosine-normalization of the co-occurrence matrix over-estimates the similarity because this matrix already contains the numerator values of the cosine (the inner products of the vectors).

The Ochiai of the co-occurrence matrix in Table 2 can be rewritten in the terms of Table 1 (the occurrence matrix) as follows:

$$Ochiai = \frac{v_1 * v_2}{\sqrt{L_2(v_1)} * \sqrt{L_2(v_2)}} \tag{7}$$

where $v_1$ is the value of the first variable in the occurrence matrix and $L_2(v_1)$is the squared norm of the vector $v_1$ in the occurrence matrix. From the rewrite in Eq. 7, it follows analytically that the Ochiai coefficients of the co-occurrence matrix are equal to the cosine similarities of the occurrence matrix as provided in Eq. 1 (*Q.e.d.*; cf. Bolton, 1991). This is true for both numerical and binary matrices.

Using SPSS, the Ochiai coefficients of the occurrence matrix are always set equal to zero or one because this measure is considered as valid only for binary matrices. If one pursues the computation numerically using Eq. 3 above for the calculation of the Ochiai coefficients, however, the cell value (V2, V3) is 1.88 (that is, larger than one), and thus invalid. In other words, the Ochiai coefficient cannot always be properly defined for the numerical case of the *occurrence* matrix. Driver & Kroeber (1932, at p. 217) formulated: "As such a coefficient, however, its validity depends on the sigmas of the values dealt with, and these cannot be ascertained for data of the kind we are dealing with.” Therefore, one should use the cosine in the case of normalizing an occurrence matrix. We will discuss the diagonal values in the case of a co-occurrence matrix below.

The bottom row of Table 4 provides the results of cosine-normalization of the *overlap matrix* (in Table 3) and the corresponding Ochiai coefficients. The cosine-normalized Table 3 significantly over-estimates the similarities, because one normalizes twice: once to generate the minimum overlap (that is, the proximity degree between the vectors which provides us with a raw (and local) similarity value.) and a second time by taking the cosine values of the resulting overlaps. Thus, one should use Ochiai coefficients also in this case.

In other words, the co-occurrence matrix of Table 2 contains the information for generating the properly normalized matrix when the diagonal values are based on multiplication of the occurrence matrix with its transposed. However, these diagonal values are often unavailable in empirical research. For example, if one queries with “a AND b” for off-diagonal values, and with only “a” or “b” for the diagonal values, these are not the squared norms of the vector ($L_2 = \sum_i x_i^2$), but the sums ($L_1 = \sum_i x_i$). In these cases, one uses *de facto* the overlap function because of the restrictive Boolean AND in the queries (Morris, 2005).

Had we used the $L_1$ norms of Table 1 {3, 6, 8} as the diagonal values in the co-occurrence matrix in Table 2, the corresponding cell (V2, V3) would again be larger than one and therefore not valid. Leaving the diagonal blank generates an error because of a division by zero. Whereas the cosine can be computed with any value on the diagonal, the Ochiai coefficient requires the diagonal values to be at least equal to the sum of the off-diagonal cells in the corresponding rows or columns of the co-occurrence matrix. Under this condition, the off-diagonal values represent subsets of the set represented on the main diagonal (Driver & Kroeber, 1932).

If the occurrence matrix is available, one can use the information contained in this matrix to construct the main diagonal as the squared norm of each vector. If the underlying occurrence matrix can be assumed to be binary, $L_1 = L_2$ and the results of using matrix multiplication or the overlap function are precisely the same. In all other cases, the diagonal values have to be equal or larger than $L_1$ of the co-occurrence matrix if one wishes to use Ochiai coefficients.

### Using Ahlgren's (2003) matrix

The co-occurrence matrix as provided by Ahlgren *et al*. (2003, Table 7, at p. 555) was reconstructed and updated by Leydesdorff & Vaughan (2006) and provided with the $L_2$ values for the main diagonal by Leydesdorff (2008, at p. 78). Note that the numbers of co-citations in Table 5 are slightly higher than those provided by Ahlgren *et al*. because the citations were retrieved at a later date (that is, Nov. 18, 2004).

**Table 5**: Author co-citation matrix of 24 information scientists in Table 7 of Ahlgren *et al*., 2003, at p. 555; main diagonal values added by Leydesdorff and Vaughan (2006; see Leydesdorff, 2008, at p. 78.)

| | | | | | | | | | | | | | | | | | | | | | | | | | |
|---|---|---|---|---|---|---|---|---|---|---|---|---|---|---|---|---|---|---|---|---|---|---|---|---|---|
| Braun | 50 | 29 | 19 | 20 | 9 | 13 | 5 | 9 | 7 | 7 | 2 | 0 | 0 | 0 | 0 | 0 | 0 | 0 | 0 | 0 | 0 | 0 | 0 | 0 | **120** |
| Schubert | 29 | 60 | 28 | 18 | 10 | 18 | 5 | 5 | 5 | 12 | 2 | 1 | 0 | 0 | 0 | 0 | 0 | 0 | 0 | 0 | 0 | 0 | 0 | 0 | **133** |
| Glanzel | 19 | 28 | 53 | 16 | 10 | 20 | 9 | 14 | 9 | 11 | 5 | 3 | 0 | 0 | 0 | 0 | 0 | 0 | 0 | 0 | 0 | 0 | 0 | 0 | **144** |
| Moed | 20 | 18 | 16 | 55 | 12 | 20 | 5 | 18 | 13 | 12 | 7 | 4 | 0 | 0 | 0 | 0 | 0 | 0 | 0 | 0 | 0 | 0 | 0 | 0 | **145** |
| Nederhof | 9 | 10 | 10 | 12 | 31 | 12 | 8 | 11 | 7 | 4 | 4 | 2 | 0 | 0 | 0 | 0 | 0 | 0 | 0 | 0 | 0 | 0 | 0 | 0 | **89** |
| Narin | 13 | 18 | 20 | 20 | 12 | 64 | 11 | 20 | 21 | 20 | 11 | 9 | 1 | 0 | 1 | 1 | 0 | 0 | 1 | 1 | 0 | 0 | 0 | 0 | **180** |
| Tijssen | 5 | 5 | 9 | 5 | 8 | 11 | 22 | 13 | 10 | 5 | 6 | 1 | 0 | 1 | 2 | 1 | 0 | 0 | 0 | 1 | 0 | 0 | 0 | 0 | **83** |
| VanRaan | 9 | 5 | 14 | 18 | 11 | 20 | 13 | 50 | 13 | 12 | 12 | 6 | 2 | 1 | 2 | 1 | 0 | 0 | 0 | 1 | 0 | 0 | 0 | 0 | **140** |
| Leydesdorff | 7 | 5 | 9 | 13 | 7 | 21 | 10 | 13 | 46 | 17 | 14 | 10 | 1 | 0 | 1 | 1 | 0 | 0 | 0 | 2 | 0 | 0 | 0 | 0 | **131** |
| Price | 7 | 12 | 11 | 12 | 4 | 20 | 5 | 12 | 17 | 54 | 10 | 9 | 1 | 1 | 1 | 1 | 0 | 0 | 2 | 0 | 1 | 0 | 1 | 2 | **129** |
| Callon | 2 | 2 | 5 | 7 | 4 | 11 | 6 | 12 | 14 | 10 | 26 | 4 | 0 | 0 | 1 | 1 | 0 | 0 | 0 | 1 | 0 | 0 | 0 | 0 | **80** |
| Cronin | 0 | 1 | 3 | 4 | 2 | 9 | 1 | 6 | 10 | 9 | 4 | 24 | 1 | 0 | 0 | 1 | 0 | 0 | 0 | 1 | 0 | 1 | 1 | 1 | **55** |
| Cooper | 0 | 0 | 0 | 0 | 0 | 1 | 0 | 2 | 1 | 1 | 0 | 1 | 30 | 15 | 5 | 12 | 5 | 10 | 7 | 2 | 0 | 2 | 1 | 1 | **66** |
| Vanrijsbergen | 0 | 0 | 0 | 0 | 0 | 0 | 1 | 1 | 0 | 1 | 0 | 0 | 15 | 30 | 7 | 17 | 5 | 13 | 5 | 3 | 1 | 0 | 1 | 1 | **71** |
| Croft | 0 | 0 | 0 | 0 | 0 | 1 | 2 | 2 | 1 | 1 | 1 | 0 | 5 | 7 | 18 | 9 | 6 | 7 | 8 | 6 | 2 | 1 | 2 | 2 | **63** |
| Robertson | 0 | 0 | 0 | 0 | 0 | 1 | 1 | 1 | 1 | 1 | 1 | 1 | 12 | 17 | 9 | 36 | 7 | 13 | 12 | 10 | 8 | 6 | 4 | 4 | **109** |
| Blair | 0 | 0 | 0 | 0 | 0 | 0 | 0 | 0 | 0 | 0 | 0 | 0 | 5 | 5 | 6 | 7 | 18 | 10 | 4 | 2 | 2 | 2 | 0 | 0 | **43** |
| Harman | 0 | 0 | 0 | 0 | 0 | 0 | 0 | 0 | 0 | 0 | 0 | 0 | 10 | 13 | 7 | 13 | 10 | 31 | 9 | 5 | 5 | 3 | 1 | 1 | **77** |
| Belkin | 0 | 0 | 0 | 0 | 0 | 1 | 0 | 0 | 0 | 2 | 0 | 0 | 7 | 5 | 8 | 12 | 4 | 9 | 36 | 9 | 9 | 10 | 14 | 10 | **100** |
| Spink | 0 | 0 | 0 | 0 | 0 | 1 | 1 | 1 | 2 | 0 | 1 | 1 | 2 | 3 | 6 | 10 | 2 | 5 | 9 | 21 | 11 | 7 | 5 | 4 | **71** |
| Fidel | 0 | 0 | 0 | 0 | 0 | 0 | 0 | 0 | 0 | 1 | 0 | 0 | 0 | 1 | 2 | 8 | 2 | 5 | 9 | 11 | 23 | 12 | 10 | 6 | **67** |
| Marchionini | 0 | 0 | 0 | 0 | 0 | 0 | 0 | 0 | 0 | 0 | 0 | 1 | 2 | 0 | 1 | 6 | 2 | 3 | 10 | 7 | 12 | 24 | 11 | 5 | **60** |
| Kuhlthau | 0 | 0 | 0 | 0 | 0 | 0 | 0 | 0 | 0 | 1 | 0 | 1 | 1 | 1 | 2 | 4 | 0 | 1 | 14 | 5 | 10 | 11 | 26 | 14 | **65** |
| Dervin | 0 | 0 | 0 | 0 | 0 | 0 | 0 | 0 | 0 | 2 | 0 | 1 | 1 | 1 | 2 | 4 | 0 | 1 | 10 | 4 | 6 | 5 | 14 | 20 | **51** |
| | **120** | **133** | **144** | **145** | **89** | **180** | **83** | **140** | **131** | **129** | **80** | **55** | **66** | **71** | **63** | **109** | **43** | **77** | **100** | **71** | **67** | **60** | **65** | **51** | **2,272** |

The values on the main diagonal were added by us on the basis of the occurrence matrix. Since this occurrence (author/document) matrix is binary, the sum in each column is equal to both the $L_1$ and $L_2$ norms of the vector. Additionally, the margin totals in Table 5 provide the total numbers of co-citations whole-number counted (excluding the main diagonal). In this case, these values are much larger than the squared norms of the corresponding vectors (on the main diagonal) because of the whole-number counting.

Since the co-citation matrix in Table 5 is derived from the asymmetrical occurrence matrix containing 279 co-citing documents as cases versus the 24 cited authors as variables, the cosine values of the occurrence matrix are (for the analytical reasons specified above) identical to the Ochiai values obtainable from the co-occurrence matrix.

Let us elaborate an example: Ahlgren *et al.* (2003, p. 558, Table 9) report a Pearson correlation between the columns (or rows) representing Van Raan and Schubert of 0.74. (The cosine value between the corresponding two columns in the co-occurrence matrix is 0.454.) However, Leydesdorff & Vaughan (2006, p. 1621, Table 3) report $r = -.131$ ($p < 0.05$) on the basis of the occurrence matrix. Thus, one can be terribly misled by using the Pearson correlation or cosine similarity based on the co-occurrence matrix. Although the co-occurrence patterns can be similar when related to the other authors in the set (sometimes considered as the global level; e.g., Colliander & Ahlgren, 2012), their local relationship is rather dissimilar. In the case of using the cosine—which runs unlike the Pearson from zero to one—the proper value of the similarity between these two vectors is 0.091, and thus consistent with the negative value of the Pearson correlation.

The highest values of the Pearson correlations reported by Ahlgren *et al*. (2003) are between Braun, Schubert, and Glänzel: 0.94 between Braun and Schubert, 0.96 between Braun and Glänzel, and 0.91 between Schubert and Glänzel. The cosine values for these cells (based on Table 5) are 0.87, 0.77, and 0.84, respectively, when the main diagonal is disregarded. The proper values, however, are 0.53, 0.37, and 0.50 using the Ochiai coefficient for the co-occurrence matrix (or equivalently the cosine for the occurrence matrix). As noted, the inflation of the cosine similarities and Pearson correlations finds its origin in the fact that the co-occurrence values are inner products of the original vectors and thus already a first step in the normalization.

**Multidimensional Scaling and Cluster Analysis**

Figure 1 shows the difference between using cosine similarity or the Ochiai coefficient for normalizing the co-occurrence matrix in Table 5 using multi-dimensional scaling in SPSS (ProxScal).[3] Whereas the left-side figure based on cosine-normalization of the co-occurrence matrix shows a strong grouping of the two subsets of authors (bibliometricians versus authors in information retrieval), it hardly shows the fine structures within each of these two groupings. The projection of the Ochiai-normalized co-occurrence matrix shows more detail about the within group structures.

[3] The variable labels are abbreviated to 10 positions in SPSS. “VANRIJSBERG” should be read as “VAN RIJSBERGEN” and “LEYDESDORF” as “LEYDESDORFF”.

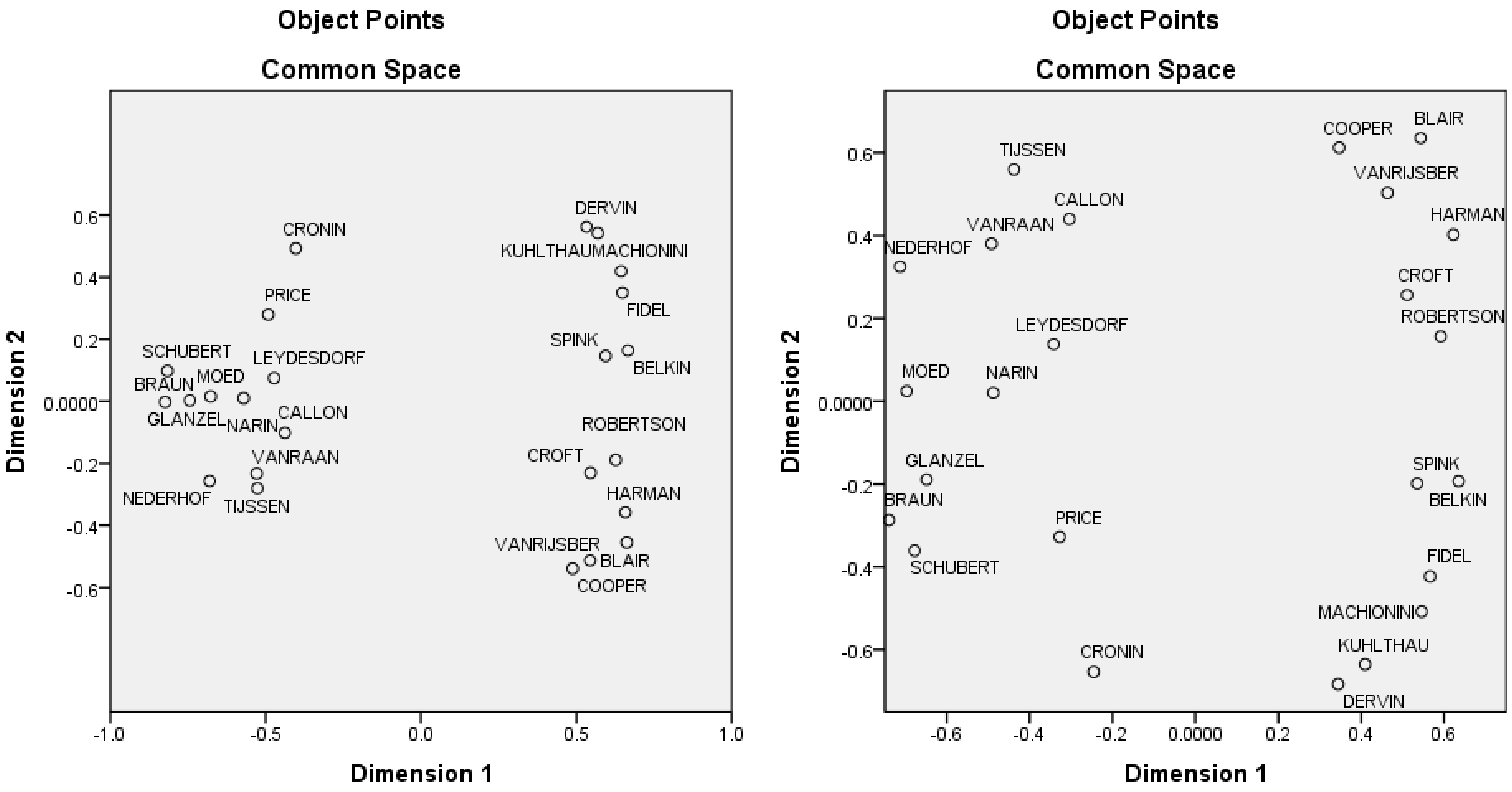

**Figure 1**: Multidimensional Scaling (PROXSCAL in SPSS) of the cosine-normalized co-occurrence matrix on the left side and the Ochiai-normalized co-occurrence matrix on the right side.

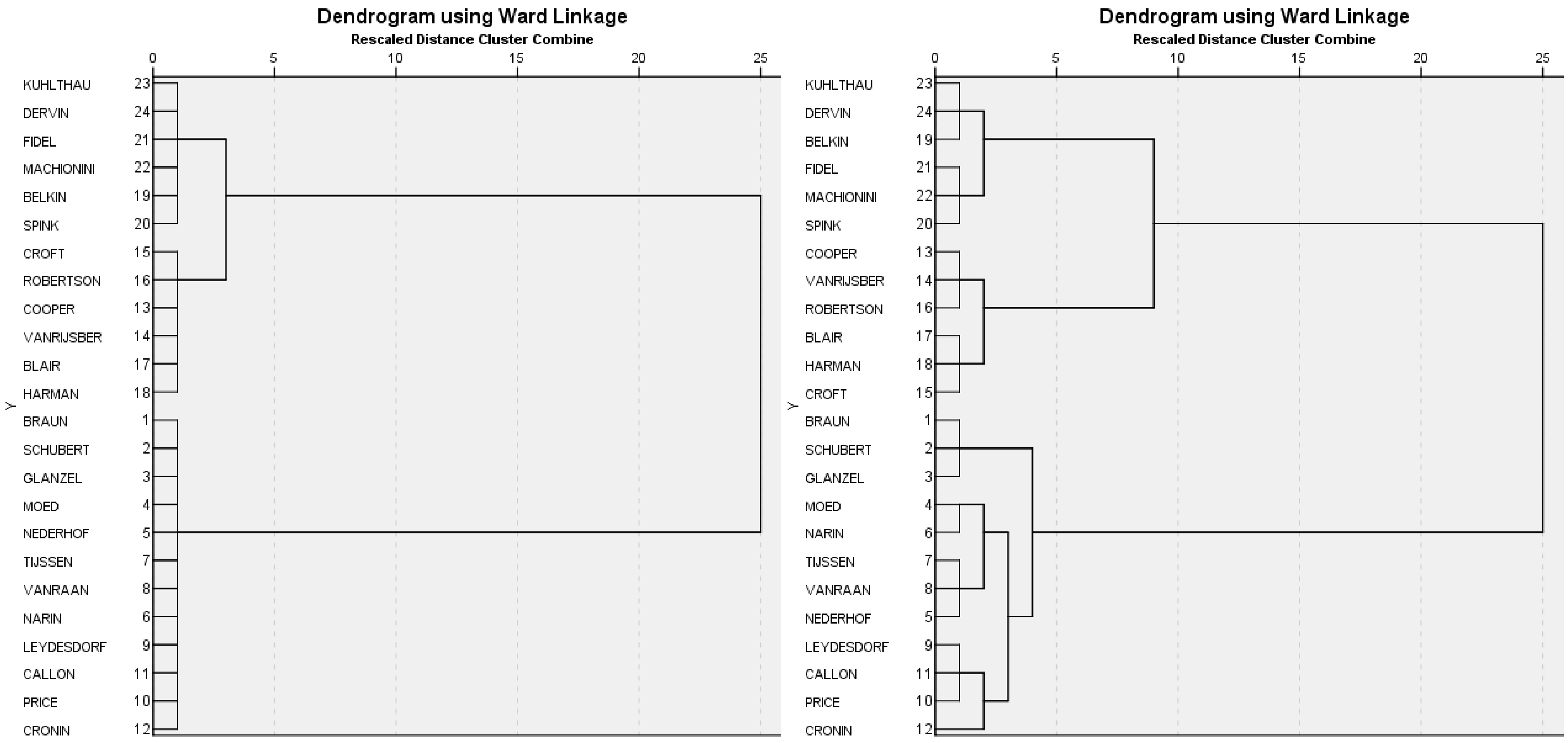

**Figure 2**: Dendograms based on Ward's clustering algorithm of Ahlgren *et al*.'s (2003) Table 7 using the cosine-normalized co-occurrence matrix on the left side and the Ochiai-normalized co-occurrence matrix on the right side.

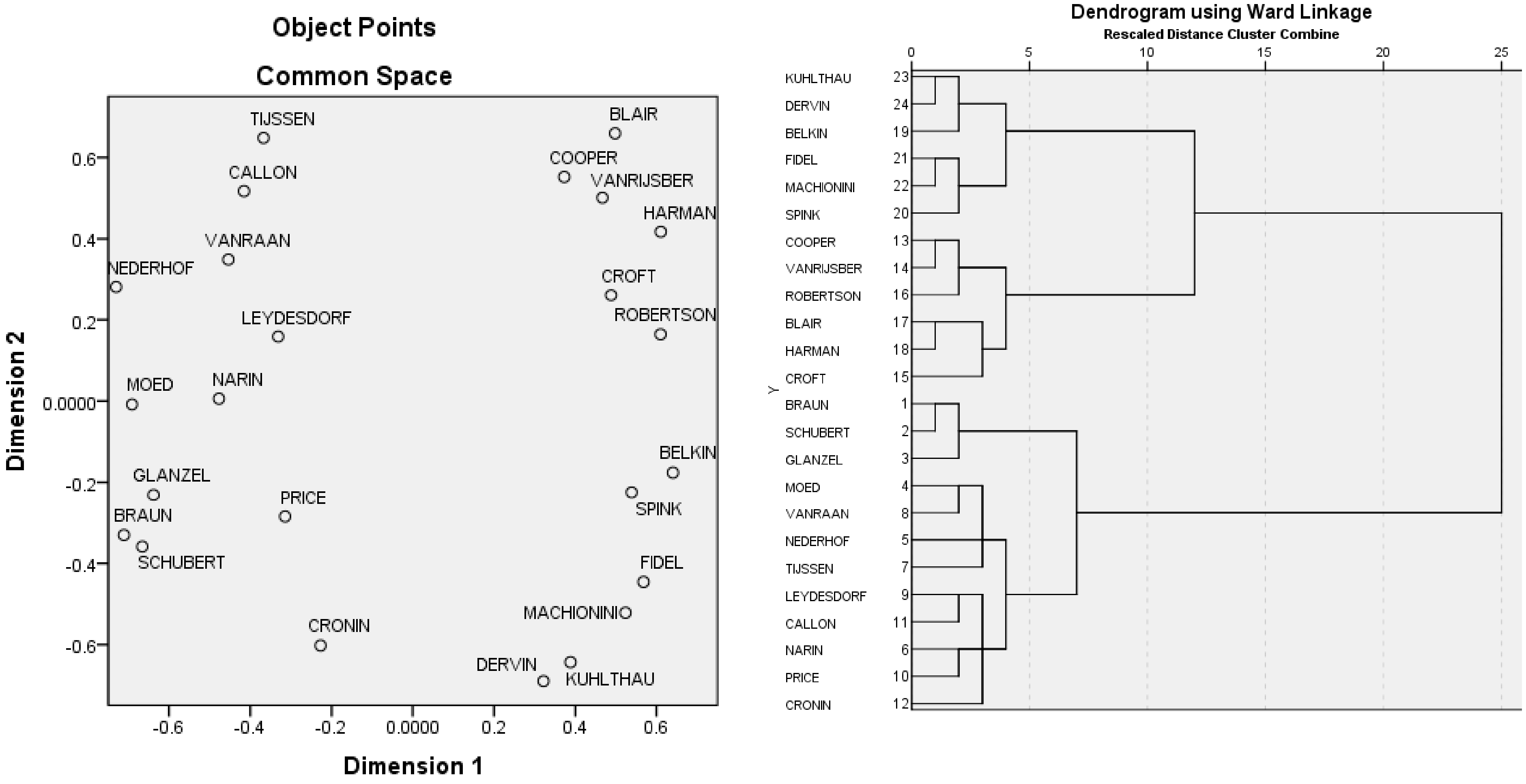

**Figure 3**: PROXSCAL and Ward's clustering of the Ochiai-normalized co-occurrence matrix, but using the sum of the off-diagonal elements for the main diagonal.

Figure 2 further refines this picture quantitatively by providing dendograms based on Ward's clustering analysis of the two matrices.[4] Whereas in the left-side picture (based on cosine-normalization) the 12 bibliometricians are all combined into a single group, the right-side dendogram (based on Ochiai normalization) shows precisely: (1) the Budapest group of Braun, Schubert and Glänzel, (2) The Leiden group, subdivided into a core group around Van Raan and including a co-citation relation between Moed and Narin, (3) a group of more theoretically oriented bibliometricians including Callon, Leydesdorff, Price, and also Cronin a bit more distantly. Similarly, a much more nuanced fine-structure is indicated among the information retrievalists. In short, the similarities in the left-side picture are over-estimated, and the Ochiai coefficient thoroughly solves the issue of properly normalizing co-occurrence matrices.

Figure 3 shows similarly the MDS and clustering solutions of the Ochiai-normalized co-occurrence matrix assuming that the occurrence matrix is not available. The main diagonal values are now provided by the sum of the off-diagonal elements for each row and column. The differences between the two MDS maps (Figures 1b and 3a) are small, but the clustering (Figure 3b) shows some differences. Narin, for example, is now placed in a cluster with Price and not with Moed and the other members of the Leiden group. The clustering in Figure 3b is more fine-grained; but the similarity is under-estimated when compared with Figure 2b. As noted, the choice of either solution depends on the research design: (1) is the occurrence matrix available for computing the squared norms of the vectors to be filled in the diagonals of the co-occurrence matrix, or (2) can it be assumed that the underlying occurrence matrix is binary.

[4] The clustering algorithm adds a normalization with the Squared Euclidean Distances by default, but this is similar for all matrices under discussion. Alternatively, one can access the normalized matrices directly using the sub-procedure MATRIX=IN(*) of CLUSTER in SPSS.

One of the referees asked to extend the analysis for a set larger than the one provided by Ahlgren *et al.* (2003). For example, Leydesdorff, Heimeriks, & Rotolo (in press) constructed a matrix with publication counts of 43 OECD nations and affiliated economies versus 10,542 journals included in JCR 2012. This matrix is an (asymmetrical) occurrence matrix. Table 6 provides the Pearson correlations, cosine values, and Spearman correlations for the first five of these countries in alphabetical order as an example.

**Table 6**: Pearson correlations, cosine values, and Spearman rank-order correlations among five nations included in the portfolio analysis of Leydesdorff, Heimeriks, and Rotolo (in press).

| | | Australia | Austria | Belgium | Canada |
|---|---|---|---|---|---|
| Austria | Pearson Correlation | 0.619 | | | |
| | Cosine | 0.635 | | | |
| | Spearman correlation | 0.425 | | | |
| Belgium | Pearson Correlation | 0.683 | 0.787 | | |
| | Cosine | 0.697 | 0.795 | | |
| | Spearman correlation | 0.526 | 0.499 | | |
| Canada | Pearson Correlation | 0.713 | 0.721 | 0.783 | |
| | Cosine | 0.727 | 0.733 | 0.793 | |
| | Spearman correlation | 0.649 | 0.440 | 0.533 | |
| Chile | Pearson Correlation | 0.379 | 0.365 | 0.386 | 0.400 |
| | Cosine | 0.391 | 0.377 | 0.398 | 0.412 |
| | Spearman correlation | 0.275 | 0.288 | 0.290 | 0.274 |

Note that the cosine is always larger than the Pearson correlation because it ranges from zero to one, whereas the Pearson correlation ranges from -1 to +1. We also added the Spearman rank correlation because this correlation has in common with the cosine that it is non-parametric.

After multiplication with the transpose one obtains the co-occurrence matrix among these 43 countries. Using the Ochiai for the co-occurrence matrix will for analytical reasons (shown above) provide us with the same values as the cosine values in Table 6. Since the argument is analytical, the equality of the cosine values of the occurrence matrix with the Ochiai values for the corresponding co-occurrence matrix holds for matrices of all sizes.

## Conclusions and discussion

We argue in this study that the proper equivalent to cosine-normalization of the occurrence matrix is Ochiai-normalization in the case of the corresponding co-occurrence matrix. We have shown both analytically and using empirical examples that the results of the two normalizations are identical. The co-occurrence matrix based on matrix multiplication conserves information about the vectors in the occurrence matrix in the values on the main diagonal.

In empirical cases, the researcher may only have retrieved a numerical co-occurrence matrix. One can then set the main diagonal, for example, to zero and accept some error in the measurement when using the cosine for the normalization, but the similarity is then overestimated. Using Ochiai coefficients for the normalization, however, the diagonal value has to be as a minimum at the sum of the off-diagonal elements in the same row or column (of this symmetrical matrix). One can consider these off-diagonal elements as subsets of the total set in each row or column. The co-occurrence matrix is then based on the overlap function (Morris, 2005; cf. Driver & Kroeber, 1932). The precise specification of the diagonal value can also be considered as a challenge for further research.

Unlike the cosine and the Ochiai coefficient, the Pearson correlation also *z*-normalizes the variation. The cosine is scale-independent, but not mass-independent, and therefore an author A with co-citations with an overall highly-cited author is more similar to this author, then the same author A with a less-cited other author irrespective of the association pattern. This caveat to the interpretation provides another option for further research and reflection. Note that Colliander & Ahlgren (2012) argued in favor of a second-order similarity matrix that would outperform the first-order one.

Furthermore, the question remains whether one should wish to normalize a co-occurrence matrix. The co-occurrence matrix itself is already normalized in terms of the inner products between the vectors and thus information-rich. In general, cosine normalization similar to Pearson normalization (and factor analysis) enables us to visualize structure in the matrix in terms of components. If one is less interested in the commonalities in the variance and more in the specificity of the various cases, one may wish to use the co-occurrence matrix *without* further normalization (e.g., Leydesdorff , Heimeriks & Rotolo, in press).

**Acknowledgements**

We thank Fuhai Leng and two anonymous referees for comments on previous drafts.